\documentclass[aps,twocolumn,showpacs,amsmath]{revtex4-1}
\usepackage{graphicx}
\usepackage{amsmath}

\begin{document}

\title{Comment on ``Anomalous Edge State in a Non-Hermitian Lattice''}

\author{Ye Xiong}
\email{xiongye@njnu.edu.cn}
\affiliation{Department of Physics and Institute of Theoretical Physics
  , Nanjing Normal University, Nanjing 210023, P. R. China\
   National Laboratory of Solid State Microstructures, Nanjing
  University, Nanjing 210093, P. R. China}

\author{Tianxiang Wang}
\affiliation{Department of Physics and Institute of Theoretical Physics
  , Nanjing Normal University, Nanjing 210023,
P. R. China}

\author{Xiaohui Wang}
\affiliation{Department of Physics and Institute of Theoretical Physics
  , Nanjing Normal University, Nanjing 210023,
P. R. China}

\author{Peiqing Tong}
\email{pqtong@njnu.edu.cn}
\affiliation{Department of Physics and Institute of Theoretical Physics
  , Nanjing Normal University, Nanjing 210023,
P. R. China \\
Jiangsu Key Laboratory for Numerical Simulation of Large
  Scale Complex Systems, Nanjing Normal University, Nanjing 210023,
P. R. China}

\maketitle

In this comment, we criticize three main conclusions of the
letter\cite{Lee2016}. We show that the concept of fractional winding
number(FWN) is factitious, Lee's conclusions on Fig. 3 are finite-size
effect and the breakdown of bulk-boundary correspondence(BBBC) cannot be
explained by ``defective''.

\begin{figure}[thb]
  \centering
  \includegraphics[width=0.42\textwidth]{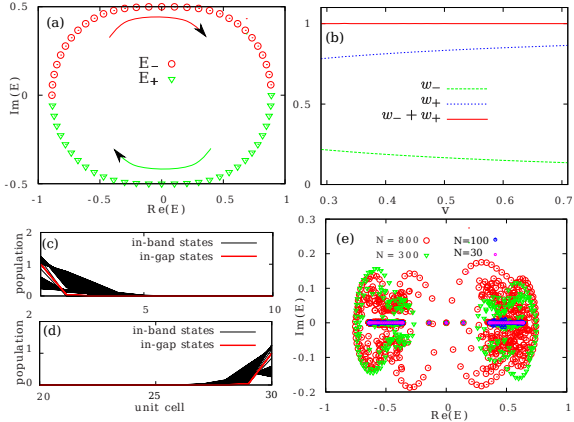}
  \caption{(color online) (a) The arrows indicates how the eigenvalues
    in the two bands evolve with wave-vector $k$. The state in the upper
    band evolves continuously to the lower band after $2\pi$ variance of
    $k$. (b) The winding numbers, $w_+$ and $w_-$ for the upper and the
    lower bands, respectively. (c) and (d) show the left and the right
    eigenvectors of a $N=30$ chain with OBC. The bulk states are changed
    from extended to localized just because the boundary condition is
    changed from periodic to open. (e) The energy spectrum for chains
    with different lengths. The spectrum are changed entirely when $N$ is
    increased. The parameters are $\gamma=1$, $r=0.5$ and $v=0.52$.}
  \label{fig1}
\end{figure}

{\it There is no FWN. ---} For a hermitian
Hamiltonian $H_h=x(k) \sigma_x + z(k) \sigma_z$, there are two
equivalent ways to calculate the winding number: one is by studying the
track of the vector $(x(k),z(k))$ in the plane and the other is by the
berry phase: $w_{\pm}=\frac{1}{\pi}\int_{k=0}^{2\pi} dk A_\pm(k)$, where
$A_{\pm}(k) =-i
\langle u_{\pm}(k) |\partial_k | u_{\pm}(k) \rangle$ is the Berry
connections for the upper $+$ and the lower $-$ bands. But in Lee's
model, the equivalence breaks down and the previous method adopted by
Lee gives a wrong answer. As Fig. \ref{fig1}(a) shows, the states are not periodic in the
Brillouin zone (BZ), e.g. the upper band states change continuously to the
lower band states after $2\pi$ variance of $k$. So the real period for
$k$ becomes $4\pi$. It is impossible to define the winding number in
{\bf one} BZ because the evolution of the states is not close in this
interval.  So FWN for each band is meaningless because it is a gauge
variant quantity. The physical quantity should be counted in the real
period $4\pi$, which will include the Berry phases of the two bands. 
So this quality must be specified by a winding number $1$.

To confirm our conclusion, we calculate $w_{\pm}$ by choosing a special
gauge: $|u_{\pm}\rangle = (1,(-r\sin(k)-i\gamma/2 \pm
E)/(r\cos(k)+v))^T$ and $\langle \langle u_\pm | = (|u_\pm \rangle)^T$.
Here $|u_{\pm}\rangle$ and $\langle \langle u_\pm |$ are the left and
the right eigenvectors and the Berry connection becomes $A_\pm= -i
 \langle \langle u_\pm | \partial_k
|u_\pm \rangle / \langle \langle u_\pm | u_\pm \rangle$ because the
model is non-hermitian. The results are presented in Fig. \ref{fig1}(b).
We find $w_{\pm}\ne 1/2$ but $w_- + w_+ =1$ as expected. By choosing
another gauge, $w_\pm$ are different.

{\it Finite-size effect. ---} In Fig. \ref{fig1}(e), we present the
spectrum for chains with open boundary condition(OBC). The length is
changed from $N=30$ to $800$. When $N=30$, our result is identical to
Lee's. But the spectrum changes entirely when $N=800$, e.g. the band gap
disappears and the spectrum becomes complex. So Lee's conclusions, such
as the PT symmetry is preserved and so on, are caused by the finite size
effect.

{\it ``Defective'' cannot explain BBBC.---} ``Defective equations'' mean
equations with less solutions. Lee used it to explain why there is only
one $E=0$ boundary state rather than two. But he further explains BBBC
with ``defective'' without any explanation. In Fig.  \ref{fig1}(c) and
(d), we show that the bulk states are changed from extended to localized
just because the boundary condition is changed. We believe that this
behavior is the real reason for BBBC. A letter supports us because it
shows that the ansatz, that most bulk states are not affected by the
boundary condition, are important in deriving the bulk boundary
correspondence(BBC)\cite{bbc}.  When the ansatz is destroyed, so as to
BBC.

\appendix
\section{The winding number}

We appreciate the author to express his version of winding number in
mathematic form. But we can prove that Eq. (3) in his third reply is
wrong because he has misunderstood the relation between the winding
number and the Brillouin zone. We will present our reasons from the
following three aspects.

Before the beginning of our discussions, we want to first present how we
will define the winding number and the difference between ours and
Lee's for the sake of clarity.

In any system, we define the berry phase as
\begin{equation}
  \gamma_B = i \oint dk \frac{\langle \langle u | \partial_k
  |u\rangle}{\langle \langle u|u\rangle}. 
  \label{eq1}
\end{equation}
The closed line integral $\oint$ stands for the closed path evolution,
which means that the wave-functions $|u\rangle$ and $\langle \langle u|$
are on a closed loop, i.e. they must return after a path in the
parameter space. We'd like to emphasize that the closed line integral
$\oint$ implicitly indicates that the loop rounds {\bf only one
time}.

The winding number, if exists, should equal to
\begin{equation}
  w=\frac{1}{\pi} \gamma_B,
  \label{eq2}
\end{equation}
irrespective to the length of the closed path. When applying this
equation to Lee's model, as the closed path of $k$ is $[0,4\pi]$, the
winding number is explicitly written as
\begin{equation}
  w=\frac{i}{\pi}  \int_0^{4\pi} dk \frac{\langle \langle u | \partial_k
|u\rangle}{\langle \langle u|u\rangle}.
\end{equation}

While Lee thinks that as the closed path goes through the Brillouin
zones twice, he divides the above winding number by $2$ to count the winding
number in {\bf one} Brillouin zone. So his version of winding number
becomes 
\begin{equation}
  w=\frac{i}{2\pi}  \int_0^{4\pi} dk \frac{\langle \langle u | \partial_k
|u\rangle}{\langle \langle u|u\rangle},
\label{eq4}
\end{equation}
which is copied from his third reply.

Now we start our discussion from {\it the first aspect}.

This is reduction to absurdity so we will adopt Lee's idea to see what
will happen.

As the divergence comes mostly from the effect of the Brillouin zone, we
take a much simple model for the sake of clarity. Our demo model is
$H(k)=\cos(k) \sigma_x + \sin(k) \sigma_y$. But we
artificially enlarge the Brillouin zone from $[0,2\pi]$ to $[0,4\pi]$.
We use BZ and BZ' to denote these two Brillouin zones. So our demo model is the
simple Hamiltonian on BZ'.

Now let's analyze the meaning of each factor in Eq. \ref{eq4}, so that
we can apply it to the demo model. First, the integral is from
$0$ to $4\pi$. This is because the states close one and {\bf only one}
loop in the $4\pi$ period in Lee's model. Second, in Lee's model the
Brillouin zone is $[0,2\pi]$. Third, there is a factor $1/2$ in
the front of the integral. This is because the integral goes through the
Brillouin zone BZ $A=2$ times. The factor in front of the integral is
$1/A$. So his equation can be written as 
\begin{equation}
  w=\frac{i}{A\pi}  \oint dk \frac{\langle \langle u | \partial_k
|u\rangle}{\langle \langle u|u\rangle}.
\end{equation}

Now let's specify the values of these factors in our demo model. First,
the closed loop is $[0,2\pi]$ as the Hamiltonian is a hermitian model.
So the integral is from $0$ to $2\pi$. Second, the Brillouin zone BZ' is
$[0,4\pi]$ as we artificially chose.  Third, $A=1/2$ as the integral
will only go through half of the Brillouin zone BZ'. Then we apply Lee's
equation and the winding number becomes  
\begin{equation}
  w=\frac{2i}{\pi}  \int_0^{2\pi} dk \frac{\langle \langle u | \partial_k
|u\rangle}{\langle \langle u|u\rangle} =2,
\end{equation}
as the integral is -i$\pi$.

We have a few remarks on the above equation. First, as the integral is
still from $0$ to $2\pi$, $w=2$ is not caused by the accumulation of the
integral in the enlarged Brillouin zone. It is only caused by the factor
$1/A$ in the front of the integral. Second, $w=2$ is wrong
because we know the winding number of the demo Hamiltonian is $w=1$. Solid state
physics tells us that enlarging the Brillouin zone is trivial
and should not introduce any modification on the physical quantities. So the
winding number must still be $1$. Third, if
we adopt our expression in Eq. \ref{eq2}, we can
get the correct winding number, $w=1$ in this case. 

The above demo model illustrates that Lee's mistake lies in his
misunderstanding on the winding number and the Brillouin zone. One may
still argue that $w=2$ seems reasonable because the Brillouin zone is
enlarged and the loop rounds twice in the new zone for the demo model.
If one adopt this explanation, then he must accept that Lee's method is
counting the winding number in one Brillouin zone, that is $[0,2\pi]$ in
his model and $[0,4\pi]$ in the demo model. In the next section, we will
come back to Lee's model and prove that this is wrong from the essential
meaning of the winding number.

{\it The second aspect.}

We want to first clarify a basic question: what is the winding number in
1-dimension? The winding number specifies a mapping from $S^1$ to $S^1$.
The previous $S^1$ is a closed path in the parameter space. For hermitian
topological insulator, this is the toroidal Brillouin zone, which is
usually denoted as $T^1$. The latter $S^1$ refers to the closed circle in
the Hilbert-space for the wave-functions.

For Lee's model, there is no such mapping from $T^1 \to S^1$ because
when $k$ changes $2\pi$, $S^1$ is not a closed loop. The mapping
restores when $k$ changes $4\pi$ and we denote such mapping as 
$2*T^1 \to S^1$. The author keeps trying to fold the mappings $2*T^1 \to
S^1$ to $T^1 \to S^1$ by dividing the previous one by a factor $2$.
But the mapping $T^1 \to S^1$ does not really
exist. Referee A has realized this point, although he/she does not
support us right now. Let's quote his/hers words here:
``the winding number is only well defined for a case, for which the
corresponding eigenvectors are on a closed loop, i.e. they must return
to an identical value after closing a path in phase space.'' 

The things
will be more clear when we throw out the concept of the Brillouin zone and consider $k$
as a parameter. In that case, our equation \ref{eq2} still works. But
Lee's equation is ill defined because he can not define the factor $1/A$ in
the front of the integral. One may still argue that Lee counts the
winding in one Brillouin zone so we cannot throw out it. This is what
we criticize in the comment and the detailed reasons are presented in the next 
section.

{\it The third aspect.}

At first, we want to clarify a few things in Lee's letter and 
those in our comment. Lee did not calculate the winding number by a
mathematic form in his Letter. His conclusion actually comes from
Fig. 2 (c) in the letter. The red solid line is the track for the
wave-functions in the upper band, $|u_+(k)\rangle$, and the red dashed
line is that for the states in the lower band,
$|u_-(k)\rangle$. We want to emphasize that {\bf these tracks are for $k\in
[0,2\pi]$}, respectively. The upper band states $|u_+\rangle$ change
to the lower band states $|u_-\rangle$ when $k$ changes $2\pi$ and a
closed loop needs totally $4\pi$ variance. Lee obtains the fractional
winding number, $w_+=0.5$ for {\bf the upper band $|u_+\rangle$} only
based on the fact that the red solid line is a half circle around $0$.
While in our comment, we calculate the winding number $w_+$ for the
upper band and find that it is a gauge variant quantity. So it is
meaningless. This is easy to be understood because the red solid line is
not a closed loop. It is only when $k$ changes $4\pi$, while the state
goes through all states in the upper band and the lower band, the loop
will close. That is why we have $w_+ + w_- =1$. This quantity is gauge
invariant and is physically meaningful.

We want to point out that Lee's equations in his third reply contradicts
with his conclusions in the Letter. If one insists that the winding
number must be defined in the Brillouin zone, he must face up to the
truth that there are two bands in this zone. Then he must inevitably
specify the two winding numbers for these two bands, respectively. (In
traditional topological insulators such as SSH model, the two winding
numbers for the two bands are identical so one uses winding number
instead of winding numbers for individual bands implicitly.) The things
are more clear when we assign the band index to Lee's equation. We will
find e.g. $w_+=\frac{i}{2\pi}  \int_0^{4\pi} dk \frac{\langle \langle u
| \partial_k |u\rangle}{\langle \langle u|u\rangle}$. Let's start the
integral from $|u(k=0)_+\rangle$. The integral on the right hand side
first goes through the upper band in the interval $[0,2\pi]$. Then it
will cover the lower band in the next $[2\pi,4\pi]$ interval. So the
right hand side of the equation includes both the contributions from
the upper band and the low band. But Lee is trying to assign it to the
upper band independently. This is of course wrong. If Lee insists that
his version of winding number is not for the upper band or the lower
band individually, but for them {\bf all}, then he must count the
contributions of the upper band and the lower band totally in the
Brillouin zone. This will give him $w=1$. So the mistake is hidden
here: This model is $4\pi$ period and the closed loop sweeps both the
upper and the lower bands. Lee notices the first property so he divide
the winding number by $2$. But he forgets the later property which
requires him to multiple $2$ because there are $2$ bands which need to
be counted together. 

Let's summary the discussions in the last paragraph because they are
very important. There are two kinds of interpretations on Lee's
equation. One is that the winding number is for the individual band. But
this will make the winding number for one band include the
contribution from the other band. The other is that the winding number
is for the two bands. Then from Fig. 2(c) in the letter, the tracks for
the two bands (the solid and the dashed red lines) totally form one loop
but not a half loop. So the winding number should be $1$ instead of
$1/2$. Our calculation also illustrates that after taking into account
the contributions from the two bands, the total winding number $w_+ +w_-
$ is $1$ but not $1/2$.

Remarks: We want to clarify why we use the integral within $[0,2\pi]$ in
our comment. This is because we are inclined to think that Lee is using
the first interpretation in his letter.
So we defined the corresponding quantities $w_\pm$ for the
two bands, respectively. But how are these quantities related to the winding
number $w$ defined in this response? In Eq. \ref{eq2}, $w=\frac{i}{\pi}
\int_0^{4\pi} dk \frac{\langle \langle u | \partial_k |u\rangle}{\langle
\langle u|u\rangle}$. Let's start the integral from $|u(k=0)_+ \rangle$.
In the interval $[0,2\pi]$, the states is in the upper band so the
integral in this interval is just $-i\pi w_+$. Then in the interval
$[2\pi,4\pi]$, the wave-functions are in the lower band. So the integral
in the later interval is $-i\pi w_-$. Now one will realize that $w=w_+ +
w_-$. This is just the argument we presented in the comment.  $w_+ + w_-=1$
is meaningful, while individuals $w_\pm$ are meaningless and the idea of
the fractional winding number is artificial. 

In summary, we have shown that Lee's equations are wrong and pointed out
where the mistake takes place. J. Phys. A 36, 2125 (2003) is right
because the berry phase there includes the contributions from both the
upper band and the lower band. But one cannot further split it half by
half and assign the two parts to the two bands, respectively.  
\bibliographystyle{apsrev}

\end{document}